# A Prudent-Precedence Concurrency Control Protocol For High Data Contention Database Environments

Weidong Xiong[1], Feng Yu[2], Mohammed Hamdi[1] and Wen-Chi Hou[1]

[1]Department of Computer Science, Southern Illinois University, Carbondale, IL 62901

[2]Department of Computer Science and Information Systems, Youngstown State University, Youngstown, OH 44555

## ABSTRACT

*In this paper, we propose a concurrency control protocol, called the Prudent-Precedence Concurrency Control (PPCC) protocol, for high data contention database environments. PPCC is prudently more aggressive in permitting more serializable schedules than two-phase locking. It maintains a restricted precedence among conflicting transactions and commits the transactions according to the serialization order established in the executions. A detailed simulation model has been constructed and extensive experiments have been conducted to evaluate the performance of the proposed approach. The results demonstrate that the proposed algorithm outperforms the two-phase locking and optimistic concurrency control in all ranges of system workload.*

## 1. INTRODUCTION

During the past few decades, there has been much research on currency control mechanisms in databases. The two-phase locking (2PL) [7], timestamping [3, 4, 13], and optimistic algorithms [10] represent three fundamentally different approaches and they have been most widely studied. Many other algorithms are developed based on these or combinations of these basic algorithms. Bernstein et al. [2] contains comprehensive discussions on various concurrency control protocols.

Optimistic concurrency controls (OCCs) have attracted a lot of attention in distributed and real-time databases [5, 6, 8, 9, 11, 12] due to its simplicity and dead-lock free nature. Transactions are allowed to proceed without hindrance until at the end - the verification phase. However, as the resource and data contention intensifies, the number of restarts can increase dramatically, and OCCs may perform much worse than 2PL [1]. As for the timestamp ordering methods, they are generally more appropriate for distributed environments with short transactions, but perform poorly otherwise [14]. 2PL and its variants have emerged as the winner in the competition of concurrency control in the conventional databases [1, 5] and have been implemented in all commercial databases.

Recent advances in wireless communication and cloud computing technology have made accesses to databases much easier and more convenient. Conventional concurrency control protocols face a stern challenge of increased data contentions, resulted from greater numbers of concurrent transactions. Although two-phase locking (2PL) [7] has been very effective in conventional applications, its conservativeness in handling conflicts can result in unnecessary blocks and aborts, and deter the transactions in high data-contention environment.

                                                                                                                              1

In this paper, we propose a concurrency control protocol, called prudent-precedence concurrency control (PPCC), for high data contention database environments. The idea comes from the observations that some conflicting transactions need not be blocked and may still be able to complete serializably. This observation leads to a design that permits higher concurrency levels than the 2PL. In this research, we design a protocol that is prudently more aggressive than 2PL, permitting some conflicting operations to proceed without blocking. We prove the correctness of the proposed protocol and perform simulations to examine its performance. The simulation results verify that the new protocol performs better than the 2PL and OCC at high data contention environments. This method is also simple and easy to implement.

The rest of this paper is organized as follows. In Section 2, we introduce the prudent-precedence concurrency control protocol. In Section 3, we report on the performance of our protocol. Conclusions are presented in Section 4.

## 2. THE PRUDENT-PRECEDENCE CONCURRENCY CONTROL

To avoid rollback and cascading rollback, hereafter we assume all protocols are strict protocols, that is, all writes are performed in the private workspaces and will not be written to the database until the transactions have committed.

### 2.1. Observations

Our idea comes from the observation that some conflicting operations need not be blocked and they may still be able to complete serializably. Therefore, we attempt to be prudently more aggressive than 2PL to see if the rationalized aggressiveness can pay off. In the following, we illustrate the observations by examples.

**Example 1.** Read-after-Write (RAW). The first few operations of transactions $T_1$ and $T_2$ are described as follows:

$T_1$: $R_1(b)$ $W_1(a)$ ...,     $T_2$: $R_2(a)$ $W_2(e)$ ...

where $R_i(x)$ denotes that transaction i reads item x, and $W_j(y)$ denotes that transaction j writes item y. Consider the following schedule:

$R_1(b)$ $W_1(a)$ $R_2(a)$ ...

There is a read-after-write (RAW) conflict on data item "a" because transaction $T_2$ tries to read "a" (i.e., $R_2(a)$) after $T_1$ writes "a" (i.e., $W_1(a)$). In 2PL, $T_2$ will be blocked until $T_1$ commits or aborts. $T_2$ can also be killed if it is blocked for too long, as it may have involved in a deadlocked situation.

If we are a little more aggressive and allow $T_2$ to read "a", $T_2$ will read the old value of "a", not the new value of "a" written by $T_1$ (i.e., $W_1(a)$), due to the strict protocol. Consequently, a read-after-write conflict, if not blocked, yields a precedence, that is, $T_2$ precedes $T_1$, denoted as $T_2 \rightarrow T_1$. We attempt to record the precedence to let the conflicting operations proceed.

**Example 2.** Write-after-Read (WAR). Consider the same transactions with a different schedule as follows.

$R_1(b)$ $R_2(a)$ $W_1(a)$ ...

Similarly, $W_1(a)$ can be allowed to proceed when it tries to write "a" after $T_2$ has read "a" ($R_2(a)$). If so, the write-after-read (WAR) conflict on item "a" produces a precedence $T_2 \rightarrow T_1$ in the strict protocol. Note that $T_2$ again reads "a" before $T_1$'s $W_1(a)$ becomes effective later in the database.

Precedence between two transactions is established when there is a read-after-write or write-after-read conflict. Note that a write-after-write conflict does not impose precedence between the transactions unless that the item is also read by one of the transactions, in which case precedence will be established through the read-after-write or the write-after-read conflicts.

Note that either in a read-after-write or write-after-read conflict, the transaction reads the item always precedes the transaction that writes that item due to the strict protocol.

## 2.2. Prudent Precedence

To allow reads to precede writes (in RAW) and writes to be preceded by reads (in WAR) without any control can yield a complex precedence graph. Detecting cycles in a complex precedence graph to avoid possible non-serializability can be quite time-consuming and defeat the purpose of the potentially added serializability. Here, we present a rule, called the Prudent Precedence Rule, to simplify the graph so that the resulting graph has no cycles and thus automatically guarantees serializability.

Let G(V, E) be the precedence graph for a set of concurrently running transactions in system, where V is a set of vertices $T_1, T_2, \ldots, T_n$, denoting the transactions in the system, and E is a set of directed edges between transactions, denoting the precedence among them. An arc is drawn from $T_i$ to $T_j$, $T_i \rightarrow T_j$, $1 \leq i, j \leq n$, $i \neq j$, if $T_i$ read an item written by $T_j$, which has not committed yet, or $T_j$ wrote an item (in its workspace) that has been read earlier by $T_i$.

Transactions in the system can be classified into 3 classes. A transaction that has not executed any conflicting operations is called an independent transaction. Once a transaction has executed its first conflicting operation, it becomes a preceding or preceded transaction, depending upon whether it precedes or is preceded by another transaction. To prevent the precedence graph from growing rampantly, once a transaction has become a preceding (or preceded) transaction, it shall remain a preceding (or a preceded) transaction for its entire lifetime.

Let $T_i$ and $T_j$ be two transactions that involve in a conflict operation. Regardless the conflict being RAW or WAR, let $T_i$ be the transaction that performs a read on the item while $T_j$ the transaction that performs a write on that item. A conflict operation is allowed to proceed only if the following rule, called the Prudent Precedence Rule, is satisfied.

**Prudent Precedence Rule:**

$T_i$ is allowed to precede $T_j$ or $T_j$ is allowed to be preceded by $T_i$ if

(i) $T_i$ has not been preceded by any transaction and

(ii) $T_j$ has not preceded any other transaction.

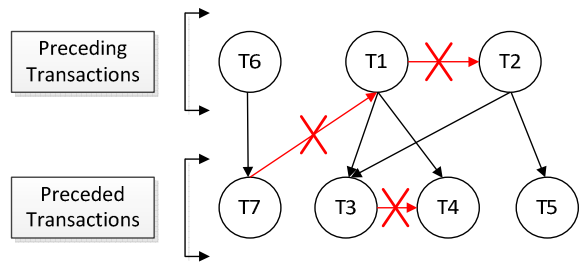

Figure 1. The Precedence Graph

We shall use Figure 1 to explain the properties of the resulting precedence graph for transactions following the Prudent Precedence Rule. It can be observed that the first condition of the rule (denoted by (i) in the rule) states that a preceded transaction cannot precede any transaction, as illustrated by the red arcs, marked with x, $T_7$ to $T_1$ and $T_3$ to $T_4$, in the figure, while the second condition (denoted (ii)) states that a preceding transaction cannot be preceded, as illustrated by the red arcs, marked with x, $T_1$ to $T_2$ and $T_7$ to $T_1$, in the figure. Since there cannot be any arcs between nodes in the same class and there is no arc from the preceded class to the preceding class, the graph cannot have a cycle.

## 2.3. Prudent Precedence Protocol

Each transaction is executed in three phases: read, wait-to-commit, and commit phases. In the read phase, transactions proceed following the precedence rule. Once a transaction finishes all its operations, it enters the wait-to-commit phase, waiting for its turn to commit following the precedence established in the read phase. Updates are written to the disk and transactions release resources in the commit phase. In the following, we describe in details each phase.

### 2.3.1. Read Phase

A transaction executing a conflict operation with another transaction will be allowed to proceed if it satisfies the prudent precedence rules; otherwise, it will be either blocked or aborted. The transaction that violates the precedence rules is hereafter called a violating transaction.
In the following, we show a situation with a violating transaction.

**Example 3.** There are three transactions. Their operations and schedule are as follows.

$T_1$: $R_1(b)$ $W_1(a)$ ...

$T_2$: $R_2(a)$ $W_2(e)$ ...

$T_3$: $R_3(e)$ ...

Schedule: $R_1(b)$ $W_1(a)$ $R_2(a)$ $W_2(e)$ $\cancel{R_3(e)}$ ...

$T_2 \rightarrow T_1$ is established when $T_2$ reads "a", and $T_2$ becomes a preceding transaction. Later when $T_3$ tries to read "e" ($R_3(e)$), the operation is suspended (denoted by $\cancel{R_3(e)}$ in the schedule) because $T_2$, a preceding transaction, cannot be preceded. Thus, $T_3$ becomes a violating transaction and needs to be blocked or aborted.

The simplest strategy to handle a violating transaction, such as $T_3$, is to abort it. Unfortunately, aborts may waste the efforts that are already spent. Therefore, we prefer blocking with the hope that the violation may later resolve and the violating transaction $T_3$ can still complete later. For example, $T_3$ is blocked, i.e., $R_3(e)$ is postponed; if $T_2$ eventually commits, then $T_3$ can resume and read the new value of "e" produced by $T_2$. The read/write with the Prudent Precedence Rule is summarized in Figure 2.

>*if there is a RAW or WAR conflict*
>
>{
>     *if the prudent precedence rule is satisfied,*
>         *proceed with the operation;*
>     *else*
>         *abort or block;*
>}

Figure 2. Read/Write with Prudent Precedence Rule

Let us elaborate on the blocking of a violating transaction a bit. By allowing a violating transaction to block, a transaction can now either be in an active (or running) state or a blocked state. Although blocking can increase the survival rate of a violating transaction, it can also hold data items accessed by the violating transaction unused for extended periods. Therefore, a time quantum must be set up to limit the amount of time a violating transaction can wait (block itself), just like the 2PL. Once the time quantum expires, the blocked (violating) transaction will be aborted to avoid building a long chain of blocked transactions.

**Theorem 1.** The precedence graph generated by transactions following the Prudent Precedence Rule is acyclic.

*Proof*: By the Prudent Precedence Rule, a preceding transaction cannot be preceded by another transaction. That is to say, in the precedence graph, there cannot be a precedence path with more than one edge. Therefore, there cannot be a cycle in the precedence graph following the Prudent Precedence Rule. As for violating transactions, they will either abort by timeouts or resume executions if the violation disappears due to the aborts or commits of the other transactions with which the transactions conflict. In either case, it does not generate any arcs that violate the Prudent Precedence Rule, and the graph remains acyclic.

### 2.3.2. Wait-to-Commit Phase

Once a transaction finishes its read phase, it enters the wait-to-commit phase, waiting for its turn to commit because transactions may finish the read phase out of the precedence order established. First, each transaction entering the wait-to-commit phase acquires exclusive locks on those items it has written in the read phase to avoid building further dependencies. Any transaction in the read phase wishes to access a locked item shall be blocked. If such a blocked transaction already preceded a wait-to-commit transaction, it shall be aborted immediately in order not to produce a circular wait, that is, wait-to-commit transactions wait for their preceding blocked transactions to complete or vice versa. Otherwise, the blocked transaction remains blocked until the locked item is unlocked. Figure 3 shows the locking when a transaction accesses a locked item.

```
/* T_i is accessing an item x */
if x is locked
{
    if x is locked by a transaction preceded by T_i
        abort T_i;
    else
        block T_i (until x is unlocked);
}
read/write with the Prudent Precedence Rule (Figure 2);
```

Figure 3. Accessing Locked Items

A transaction can proceed to the commit phase if no transactions, either in the read or the wait-to-commit phase, precede it. Otherwise, it has to wait until all its preceding transactions commit.

### 2.3.3. Commit Phase

As soon as a transaction enters the commit phase, it flushes updated items to the database, releases the exclusive locks on data items obtained in the wait-to-commit phase, and also releases transactions blocked by it due to violations of the precedence rule. Figure 4 summarizes the wait-to-commit and the commit phases.

International Journal of Database Management Systems ( IJDMS ) Vol.8, No.5, October 2016

**Example 4.** Suppose that we have the following transactions, $T_1$, $T_2$:

$T_1$: $R_1(a)$, $R_1(b)$

$T_2$: $R_2(b)$, $W_2(a)$, $W_2(b)$

Assume that the following is the schedule:

$R_1(a)$, $R_2(b)$, $W_2(a)$, $W_2(b)$, [$wc_2$], ~~$R_1(b)$~~ $abort_1$, $wc_2$, $c_2$

When $T_2$ writes "a" ($W_2(a)$), $T_1 \rightarrow T_2$ is established, due to an earlier $R_1(a)$. So, when $T_2$ reaches its wait-to-commit phase, denoted by $wc_2$, it locks both "a" and "b". However, $T_2$ has to wait until $T_1$ has committed or aborted, denoted by [$wc_2$], due to the established precedence $T_1 \rightarrow T_2$. Later, when $T_1$ tries to read "b", it is aborted, as indicated by ~~$R_1(b)$~~ and $abort_1$, because "b" is locked by $T_2$, as stipulated in Figure 3. Now no transaction is ahead of $T_2$, so it can finish its wait phase ($wc_2$) and commits ($c_2$).

>  /\* when a trans. $T_i$ reaches its wait-to-commit phase \*/
> **Wait-to-Commit Phase:**
>   Lock items written by $T_i$;
>   $T_i$ waits until all preceding transactions have committed or aborted;
> **Commit Phase:**
>   Flush updated items to database;
>   Release locks;
>   Release transactions blocked by $T_i$;

Figure 4. Wait-to-Commit and Commit Phases

## 2.4. Serializability

A history is a partial order of the operations that represents the execution of a set of transactions [5]. Let H denote a history. The serialization graph for H, denoted by $SG_H$, is a directed graph whose nodes are committed transactions in H and whose edges are $T_i \rightarrow T_j$ ($i \neq j$) if there exists a $T_i$'s operation precedes and conflicts with a $T_j$'s operation in H. To prove that a history H is serializable, we only have to prove that $SG_H$ is acyclic.

**Theorem 2**: Every history generated by the Prudent Precedence Protocol is serializable.
*Proof*: The precedence graph is acyclic as proved in Theorem 1. The wait-to-commit phase enforces the order established in the precedence graph to commit. So, the serialization graph has no cycle and is serializable.

## 3. SIMULATION RESULTS

This section reports the performance evaluation of 2PL, OCC, and the Prudent Precedence Concurrency Control (PPCC) by simulations.

## 3.1. Simulation Model

We have implemented 2PL, OCC, and PPCC in a simulation model that is similar to [1]. Each transaction has a randomized sequence of read and write operations, with each of them separated by a random period of a CPU burst of $15 \pm 5$ time units on average. The randomized disk access time is $35 \pm 10$. All writes are performed on items that have already been read in the same transactions. All writes are stored in private work space and will only be written to the database after commits following the strict protocol.

## 3.2. Parameter Settings

Our goal is to observe the performance of the algorithms under data and resource contentions. The write operations cause conflicts and thus data contentions. Therefore, we shall experiment with different write probabilities, 20% (moderate), and 50% (the highest), to observe how the two algorithms adapt to conflicts. Other factors that affect the data contentions are database sizes and transaction sizes. Therefore, two database sizes of 100 and 500 items, and two transaction sizes of averaged 8 and 16 operations will be used in the simulation.

To observe the effect of the resource contention, we report the results of simulations in which one with 4 CPUs and 8 Disks (denoted as 4/8 in Table 1) and the other 16 CPUs and 32 disks (16/32). Table 1 summarizes the base parameter settings that underline the simulation.

Table 1. Parameter Settings

| Database size | 100, 500 items |
| Average transaction size | 8 ± 4, 16 ± 4 operations |
| Write probability | 20%, 50% |
| Num. of CPUs/Disks | 4/8, 16/32 |
| CPU burst | 15 ± 5 time units |
| I/O access time | 35 ± 10 time units |

Transactions may be blocked in 2PL and PPCC to avoid generating cycles in the precedence graphs. Blocked transactions are aborted if they have been blocked longer than specified periods. We have experimented with several block periods and select the best ones to use in the simulations.

The primary performance metric is the system throughput, which is the number of transactions committed during the period of the simulation.

## 3.2. Experimental Results

In this section, we report the simulation results on the three protocols based on the above setups.

### 3.2.1. Data Contention

As mentioned earlier, the data contention is mainly caused by the write operations. If transactions have no writes, there will be no conflicts and all three protocols will have identical performance. Given the same write probability, the greater the transaction sizes, the greater the numbers of write operations are in the system, and thus the higher the data contentions are. On the other hand, given the same number of write operations, the smaller the database size, the greater the chance of conflicts, and thus the higher the data contentions are. Here, we will see how these factors affect the performance of the two protocols.

We experimented with two database sizes, 100 items and 500 items, and two transaction sizes, averaged 8 and 16 operations in each transaction. The experimental results in this subsection were obtained with the setup of 4 CPUs and 8 Disks. The simulation time for each experiment is 100,000 time units.

- Write probability 0.2

Given the write probability 0.2, each transaction has on average one write operation for every four reads.

Figures 5 and 6 show the performance for transactions with averaged 8 (8 ± 4) operations for two databases of sizes 500 (Figure 5) and 100 (Figure 6). As observed, as the level of concurrency increased initially, the throughput increased. At low concurrency levels, all protocols had similar throughputs because there were few conflicts. But as the concurrency level increased further, conflicts or data contention intensified and the increase in throughput slowed down a bit. After a particular point, each protocol reached its peak performance and started to drop, known as *thrashing*.

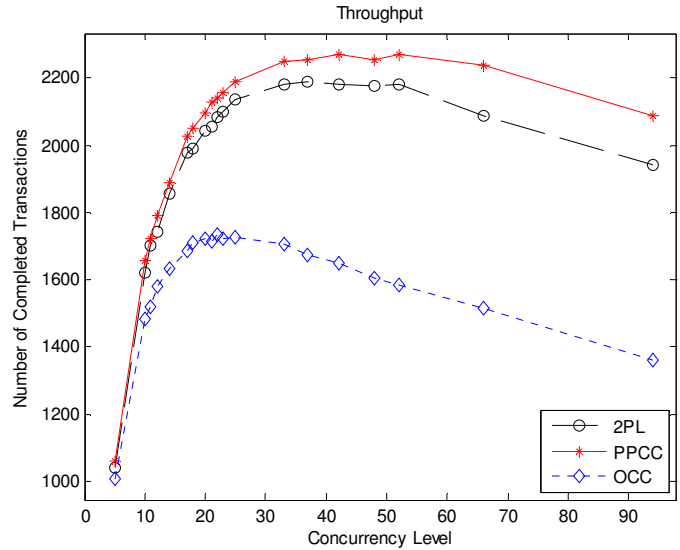

Figure 5. Write Probability 0.2, Transaction Size 8, DB Size 500

For database size 500 (Figure 5), the highest numbers of transactions completed in the given 100,000 time unit period were 2,271 for PPCC, 2,189 for 2PL, and 1,733 for OCC, that is, a 3.75% and 31.04% improvements over 2PL and OCC, respectively.

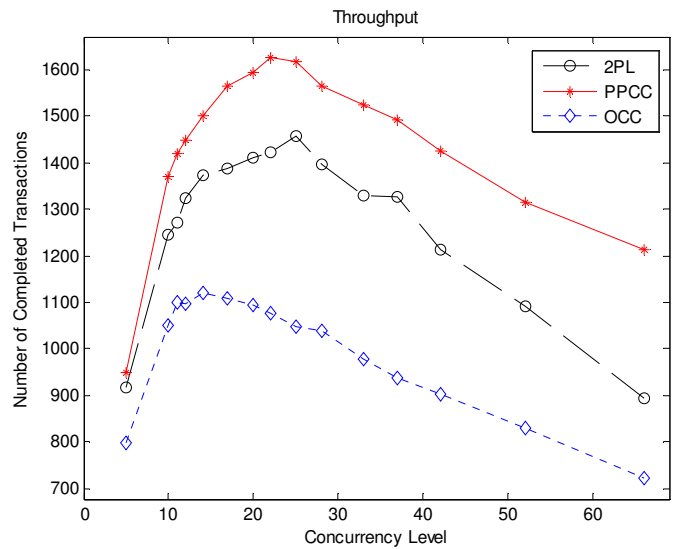

Figure 6. Write Probability 0.2, Transaction Size 8, DB Size 100

In Figure 6, the database size was reduced to 100 items to observe the performance of these protocols in a high data contention environment. The highest numbers of completed transactions were 1,625, 1,456, and 1,121 for PPCC, 2PL, and OCC, respectively, i.e., an 11.61% and 44.96% higher throughputs than 2PL and OCC. This indicates that PPCC is more effective in high data contention environments than in low data contention environments, which is exactly the purpose that we design the PPCC for.

Now, we increase the average number of operations in each transaction to 16 while maintaining the same write probability 0.2. Figures 7 and 8 show the results.

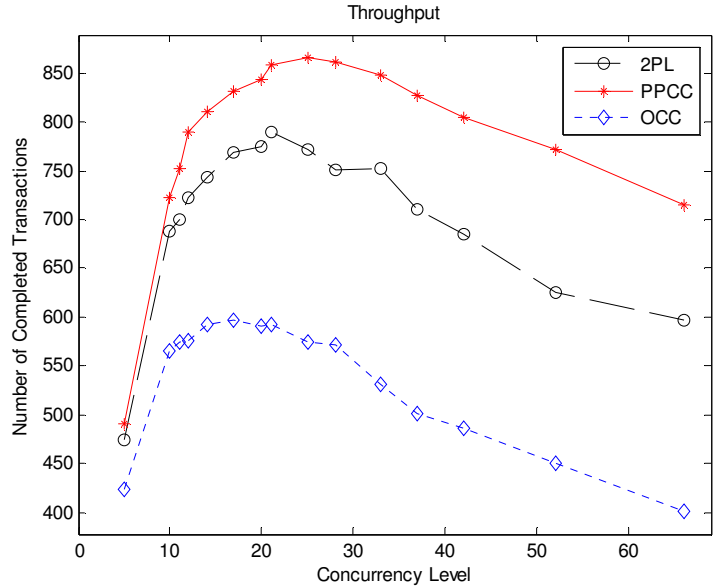

Figure 7. Write Probability 0.2, Transaction Size 16, DB Size 500

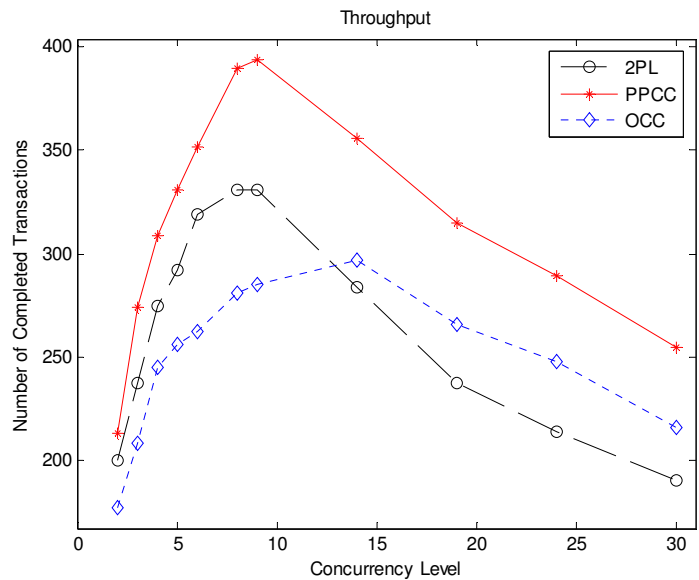

Figure 8. Write Probability 0.2, Transaction Size 16, DB Size 100

For database size 500 (Figure 7), the highest throughput obtained by PPCC was 866, while 2PL peaked at 789 and OCC at 597. PPCC had a 9.76% and 45.06% higher throughputs than 2PL and OCC. As for database size 100 (Figure 8), the highest throughputs obtained were 394, 331, and 297 for PPCC, 2PL and OCC, respectively. PPCC had a 19.03% and 32.66% higher throughputs than 2PL and OCC.

In general, as the data contention intensifies, PPCC has greater improvements over 2PL and OCC in performance.

- Write probability 0.5

With the write probability 0.5, every item read in a transaction is later written too in that transaction. Figures 9 and 10 show the throughput of the three protocols with the average number of operations set to 8 per transaction.

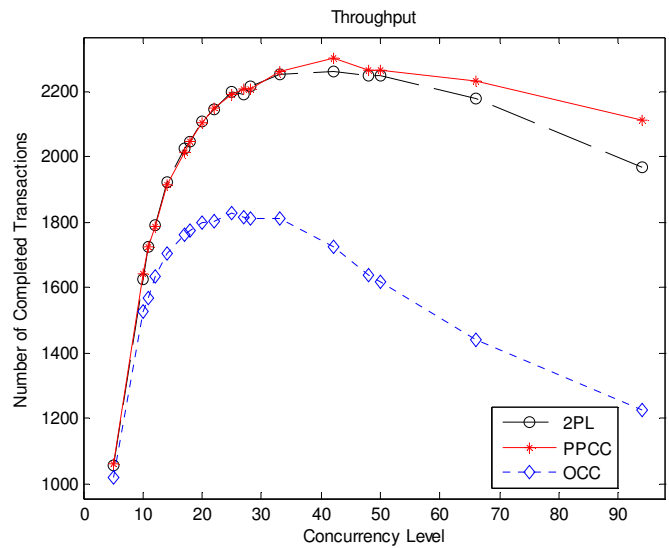

Figure 9. Write Probability 0.5, Transaction Size 8, DB Size 500

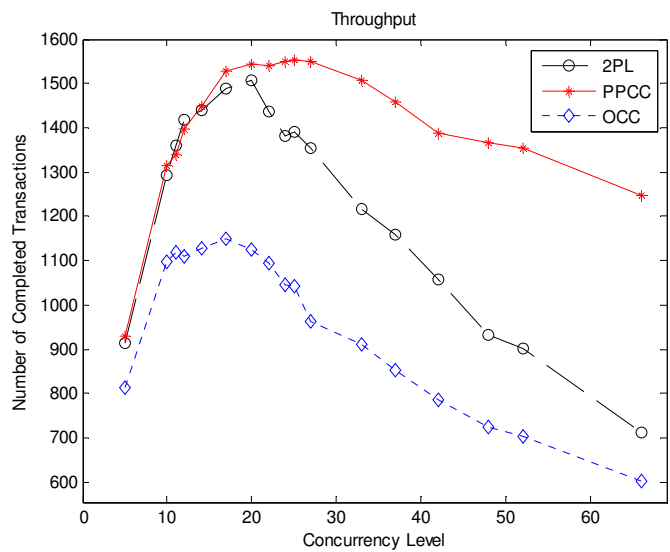

Figure 10. Write Probability 0.5, Transaction Size 8, DB Size 100

The highest numbers of transactions completed during the simulation period (Figure 9) were 2,301 for PPCC, 2,259 for 2PL, and 1,825 for OCC for database size 500, a slight improvement over 2PL(1.86%), but a much larger improvement over OCC (26.08%). As the database size decreased to 100 (Figure 10), the highest numbers of completed transactions were 1,553, 1,506, and 1,148 for PPCC, 2PL, and OCC, respectively, that is, a 3.12% and 35.28% higher throughput than 2PL and OCC, due to the higher data contentions.

Figures 11 and 12 show the throughputs of the three protocols with the number of operations per transaction increased to 16.

The highest numbers of transactions completed during the simulation period (Figure 11) were 796 for PPCC, 780 for 2PL, and 562 for OCC for database size 500, a 2.05% and 41.64% improvements over 2PL and OCC. As the database size decreased to 100 (Figure 12), the highest numbers of completed transactions were 343, 303, 283 for PPCC, 2PL, and OCC, respectively, that is, a 13.2% and 21.20% higher throughputs than 2PL and OCC.

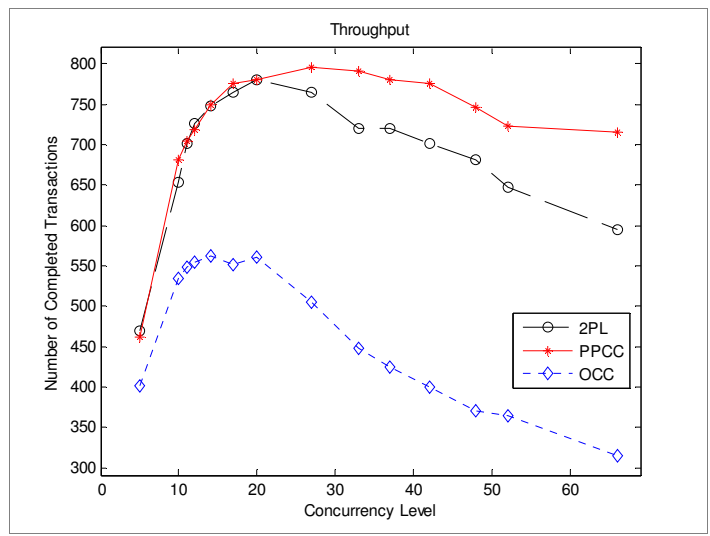

Figure 11. Write Probability 0.5, Transaction Size 16, DB Size 500

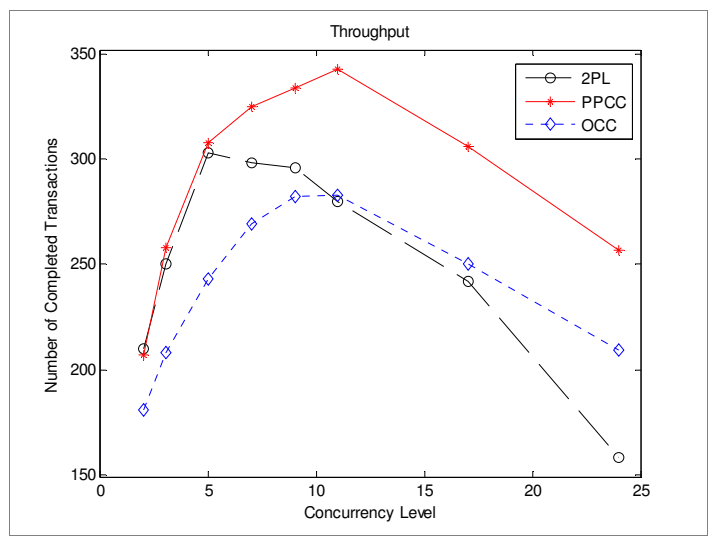

Figure 12. Write Probability 0.5, Transaction Size 16, DB Size 100

In very high data contention environments, few transactions can succeed, as illustrated in Figure 12. This indicates that there is still room for improvement in designing a more aggressive protocol that allows more concurrent schedule to complete serializably.

### 3.2.2. Resource Contention

As the hardware cost becomes cheaper and cheaper, a database can afford more CPUs and disks. Here, we examine how resource abundance can affect the three protocols. The experimental results in this subsection were obtained with the setup of 16 CPUs and 32 Disks. The simulation time for each experiment is 100,000 time units. For simplicity, we present only the cases with transaction size equal to 8.

- Write probability 0.2

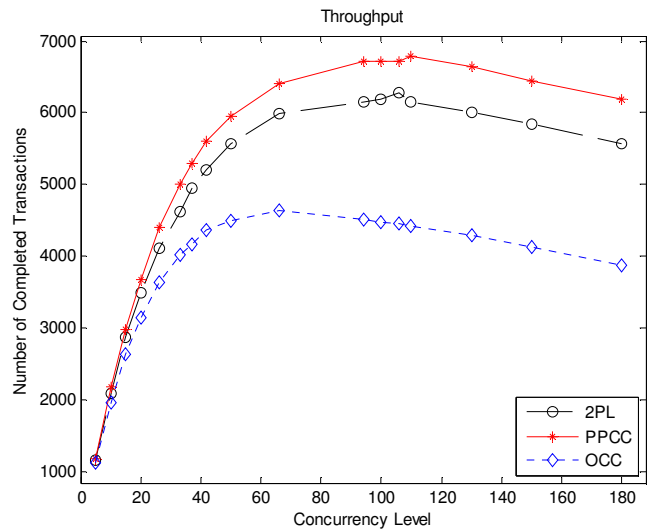

Figure 13. Write Probability 0.2, Transaction Size 8, DB Size 500

For database size 500 (Figure 13), the highest numbers of transactions completed in the given 100,000 time unit period were 6,793 for PPCC, 6,287 for 2PL, and 4,650 for OCC, that is an 8.05% and 46.09% improvements over 2PL and OCC, respectively.

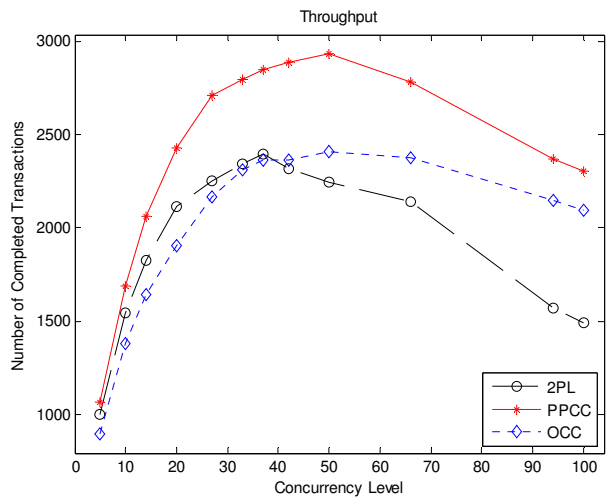

Figure 14. Write Probability 0.2, Transaction Tize 8, DB Size 100

In Figure 14, the database size was reduced to 100 items to observe the performance of the two protocols in a higher data contention environment. The highest numbers of completed transactions were 2,936, 2,400, and 2,413 for PPCC, 2PL, and OCC, respectively, i.e., a 22.33% and 21.67% higher throughputs than 2PL and OCC. This again indicates that PPCC is more effective in higher data contention environments than in lower data contention environments.

- Write probability 0.5

The highest numbers of transactions completed during the simulation period (Figure 15) were 6,659 for PPCC, 6,267 for 2PL, and 4,818 for OCC for database size 500, a 6.25% and 38.21% improvements over 2PL and OCC, respectively. As the database size decreased to 100 (Figure 16), the highest numbers of completed transactions were 2,784, 2,227, and 2,459 for PPCC, 2PL, and OCC, respectively, that is, a 25.01% and 13.22% higher throughput than 2PL and OCC, due to the higher data contentions.

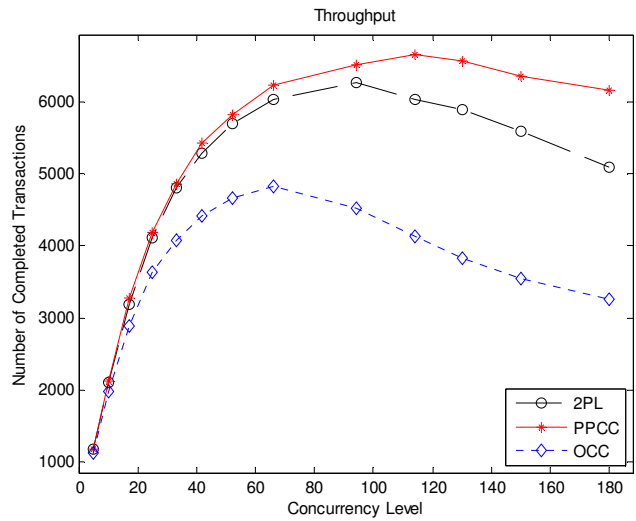

Figure 15. Write Probability 0.5, Transaction Size 8, DB Size 500

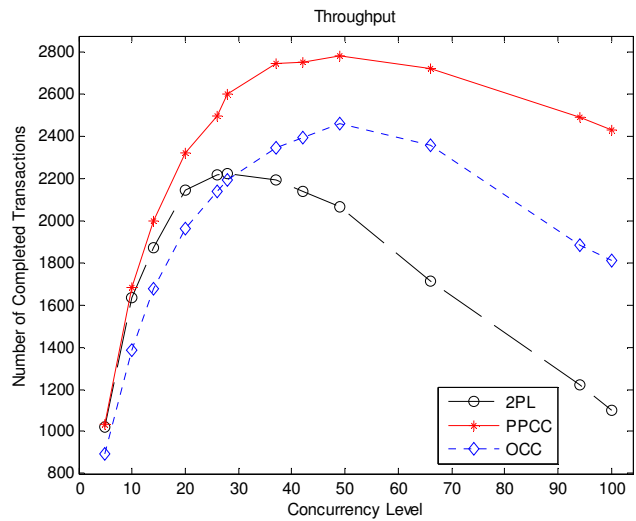

Figure 16. Write Probability 0.5, Transaction Size 8, DB Size 100

It is noted in Figures 14 and 16, OCC outperformed 2PL when the database sizes are 100. This is because restarts (or aborts) in OCC become more beneficial than long waits in 2PL when resources are abundant. Since PPCC allows more schedules to complete, it alleviates the adverse effect of long waits and performs better than both OCC and 2PL.

## 4. CONCLUSIONS

The proposed protocol can resolve conflicts successfully to a certain degree. It performed better than 2PL and OCC in all the tested situations. It has the best performance when conflicts are not extremely severe, for example, in situations where transactions are not very long and write probabilities are not too high. Further research is still needed to allow more concurrent serializable schedules to complete while keeping the protocols simple.